\documentclass[11pt,a4paper]{article}
\pdfoutput=1
\usepackage{jcappub}

\usepackage{graphicx}
\usepackage{amsmath}
\usepackage{epsfig,multicol,bbm}

\newcommand{\be}{\begin{equation}}
\newcommand{\ee}{\end{equation}}
\newcommand{\bea}{\begin{eqnarray}}
\newcommand{\eea}{\end{eqnarray}}

\newcommand{\ti}{\times}

\newcommand{\mc}{\mathcal}

\newcommand\fverb{\setbox\fverbbox=\hbox\bgroup\verb}
\newcommand\fverbdo{\egroup\medskip\noindent
			\fbox{\unhbox\fverbbox}\ }
\newcommand\fverbit{\egroup\item[\fbox{\unhbox\fverbbox}]}
\newbox\fverbbox

\begin{document}

\title{ALP Conversion and the Soft X-ray Excess in the Outskirts of the Coma Cluster}

\abstract{It was recently found that the soft X-ray excess in the center of the Coma cluster can be fitted by conversion of axion-like-particles (ALPs) of a cosmic axion background (CAB) to photons. We extend this analysis to the outskirts of Coma, including regions up to 5 Mpc from the center of the cluster. We extract the excess soft X-ray flux from ROSAT All-Sky Survey data and compare it to the expected flux from ALP to photon conversion of a CAB.  The soft X-ray excess both in the center and the outskirts of Coma can be simultaneously fitted by ALP to photon conversion of a CAB. Given the uncertainties of the cluster magnetic field in the outskirts we constrain the parameter space of the CAB. In particular, an upper limit on the CAB mean energy and a range of allowed ALP-photon couplings are derived.}

\author[a]{David Kraljic,}
\author[a]{Markus Rummel,}
\author[a]{Joseph P. Conlon}

\affiliation[a]{Rudolf Peierls Centre for Theoretical Physics, University of Oxford,\\ 1 Keble Road, Oxford, OX1 3NP, United Kingdom}
\emailAdd{j.conlon1@physics.ox.ac.uk}\emailAdd{David.Kraljic@physics.ox.ac.uk}\emailAdd{Markus.Rummel@physics.ox.ac.uk}

\maketitle

\section{Introduction}

This paper shows how the measured soft X-ray excess in the outer regions of the Coma cluster can be explained by a conversion of
axions - more strictly, axion-like particles (ALPs) - originating from a homogeneous Cosmic Axion Background (CAB) into photons.
This extends the analysis of~\cite{2013arXiv1312.3947A} for the central region of Coma.

Galaxy clusters - gravitationally bound assemblies of 100 - 1000 galaxies - are the largest virialised objects in the universe. An important component of galaxy clusters is the intracluster medium (ICM) which is a hot ionised plasma with multi-keV temperatures and characteristic electron number density $n_e \sim 10^{-3} - 10^{-5}\, \text{cm}^{-3}$. The ICM represents
approximately $10 \%$ of the overall mass of the cluster ($\sim 90 \%$ being dark matter), but is
 by far the dominant component of baryonic matter, comprising approximately $90 \%$ of baryonic matter within the cluster. The ICM is visible through
 its diffuse X-ray emission from thermal bremsstrahlung from the hot plasma.

For many clusters however,
thermal bremsstrahlung of the ICM fails to account for all of the observed X-ray flux
in the soft component of the X-ray spectrum ($E_{\gamma} \sim \mathcal{O}(0.2)$ keV). This soft excess was first discovered in~\cite{1996ApJ...458L...5L,1996Sci...274.1335L,1996Sci...274.1338B} for the Coma and Virgo clusters, using observations from the EUVE satellite, before being
  subsequently confirmed with ROSAT, see e.g.,~\cite{2002ApJ...576..688B,Durret}. In a study of 39 clusters with ROSAT, the authors of \cite{2002ApJ...576..688B}
  found a soft excess in 20 clusters at 90$\%$ confidence level, including some cases of extremely high statistical significance. There have been fewer studies of the soft excess
  with newer instruments such as XMM-Newton, Chandra or Suzaku, as these instruments are actually suboptimal for this purpose
  compared to ROSAT. This is due to smaller fields of view, higher internal backgrounds
  and reduced sensitivity to the softest X-rays. The soft excess is reviewed in \cite{Durret} and \cite{2013arXiv1312.3947A}.

The Coma cluster offers optimal conditions for the study of the soft excess.
The cluster is relatively close $(z=0.024)$, and near the galactic north pole with a low absorbing hydrogen column density.
The soft excess for Coma is also particularly pronounced and extends to large distances - it has been detected up to $2.6$ Mpc in~\cite{2003ApJ...585..722B} and more recently up to $5$ Mpc in~\cite{Bonamente:2009ns}.

The potential astrophysical explanations of the soft excess
are either an additional `warm' gas with temperatures of $\mathcal{O}(0.1)$ keV, coexisting with the hot ICM, or inverse Compton scattering of relativistic electrons with CMB photons. Both explanations have observational problems. The presence of a warm gas of $\mathcal{O}(0.1)$ keV temperature necessarily implies associated thermal emission lines, particularly ${\rm O}_{\text{VI}}$ and ${\rm O}_{\text{VII}}$, should be detected. However searches for these have proved null~\cite{1996ApJ...469L..77D,2001ApJ...550L..25D}.
Furthermore, in the cluster centre such warm gases have cooling times much shorter than the age of the cluster \cite{1996Sci...271.1244F,1997Sci...275...48F}.
The inverse Compton explanation involves $E \sim \mc{O}(300)\, \hbox{MeV}$ electrons, and so requires associated gamma ray production from non-thermal bremsstrahlung.
This then suffers both from null detection of these gamma rays from clusters~\cite{2010ApJ...717L..71A,2012JCAP...07..017A,2012MNRAS.427.1651H,2013arXiv1308.5654A,2013A&A...560A..64H,2014MNRAS.440..663Z}, and
 also a potentially overly large level of synchrotron emission~\cite{2002ApJ...566..794T}.

While it is possible that either the above or other astrophysical explanations will be understood in the future to explain the soft excess,
it is worth discussing a possible cosmological origin of this phenomenon. In particular, it was proposed in~\cite{Conlon:2013txa} that the soft excess could
originate from a CAB whose ALPs transform to photons in the large scale magnetic fields of galaxy clusters.
Such a CAB typically arises in string-theoretic descriptions of the very early universe~\cite{Conlon:2013isa}, being produced by hidden sector decays of moduli
to ALPs.\footnote{In more detail:
 when compactifying string theory to 4 dimensions, an $\mathcal{O}(100)$ number of gravitationally coupled
 scalar fields $\phi$, moduli, appear in the 4D effective action. During inflation these moduli are
 displaced from their minimum by the large inflationary energy.
 After inflation, the displaced moduli fields oscillate around their minima, behaving as matter with an
 energy density scaling as $a(t)^{-3}$ whereas radiation scales as $a(t)^{-4}$. As the universe expands the moduli then come to dominate
 the energy density. As the moduli decay rates are $\Gamma_\phi \sim m_\phi^3/M_{\text{P}}^2$, it is the lightest modulus that survives the longest and is
 responsible for reheating~\cite{Cicoli:2012aq,Higaki:2012ar, Higaki2013}. In the specific framework of the Large Volume Scenario (LVS) in type IIB string theory~\cite{Balasubramanian:2004uy, Balasubramanian:2005zx}, the branching ratios of the lightest modulus are studied in~\cite{Cicoli:2012aq,Higaki:2012ar, Angus2013, Hebecker:2014gka,Angus:2014bia}, and generically, there is a significant branching ratio to a light ALP. The ALPs produced by the
  modulus decay propagate freely to the present day, forming a homogeneous and isotropic CAB with a non-thermal spectrum set
  by the time evolution of the scale factor during modulus decay. For moduli masses in the range $\mathcal{O}(10^4 - 10^8)$ GeV, typical in string constructions,
  the energies today of the CAB overlap significantly with the $0.1 -1$ keV soft X-ray band.}

The CAB can be regarded as a contribution to dark radiation, i.e., an additional very weakly coupled species acting as additional radiation. This can be expressed in terms of the effective number of neutrinos $N_{\text{eff}}=3.046+\Delta N_{\text{eff}}$ where the first term gives the contribution of the three neutrino species corrected for thermal decoupling. Currently there are observational hints on the existence of dark radiation at the $1-3\,\sigma$-level from CMB measurements~\cite{Ade:2013zuv} and BBN~\cite{Cooke:2013cba}.

The ALPs $a$ of the CAB interact with photons via an operator
\begin{equation}
\frac{a}{M} \, \boldsymbol{E}\cdot \boldsymbol{B}\,,
\end{equation}
where $\boldsymbol{E}$ and $\boldsymbol{B}$ are the electric and magnetic fields, and $M$ is the suppression scale of the ALP-photon coupling. We stress here that strictly this paper refers only to axion-like particles; in particular, the particle $a$ we are discussing is not the QCD axion: strictly it is an ALP. Such ALPs convert to photons in coherent magnetic fields, see e.g.,~\cite{Raffelt:1987im}. Galaxy clusters are extended over $\mathcal{O}(\text{Mpc})$ distances, supporting $\mc{O}( \mu G) $ magnetic field strengths with kpc coherence scales. This combination makes clusters very efficient in converting ALPs to photons, with conversion probabilities for $M \sim 10^{13}\, \hbox{GeV}$ reaching up to $\mathcal{O}(10^{-3})$ depending on the precise details of the cluster magnetic field~\cite{Conlon:2013txa}. This fact implies that were a CAB to exist, the first place it would be observed would be as a soft excess from galaxy clusters.

As a first analysis of this possibility, the authors of~\cite{2013arXiv1312.3947A} studied ALP to photon conversion in the central $\text{Mpc}^3$ of the Coma cluster.
The magnetic field structure in this region is known moderately well from Faraday rotation measures~\cite{2010A&A...513A..30B}.
The authors of~\cite{2013arXiv1312.3947A} were able to show that for cluster magnetic field spectra consistent with Faraday rotation measurements,
the magnitude and morphology of the soft excess can be explained by ALP-photon conversion.
They also identified the resulting allowed parameter space in terms of the photon to ALP coupling $M$ and the CAB mean energy $\langle E_{CAB} \rangle$.

\begin{figure}[h!]
\centering
\includegraphics[width=0.9\linewidth]{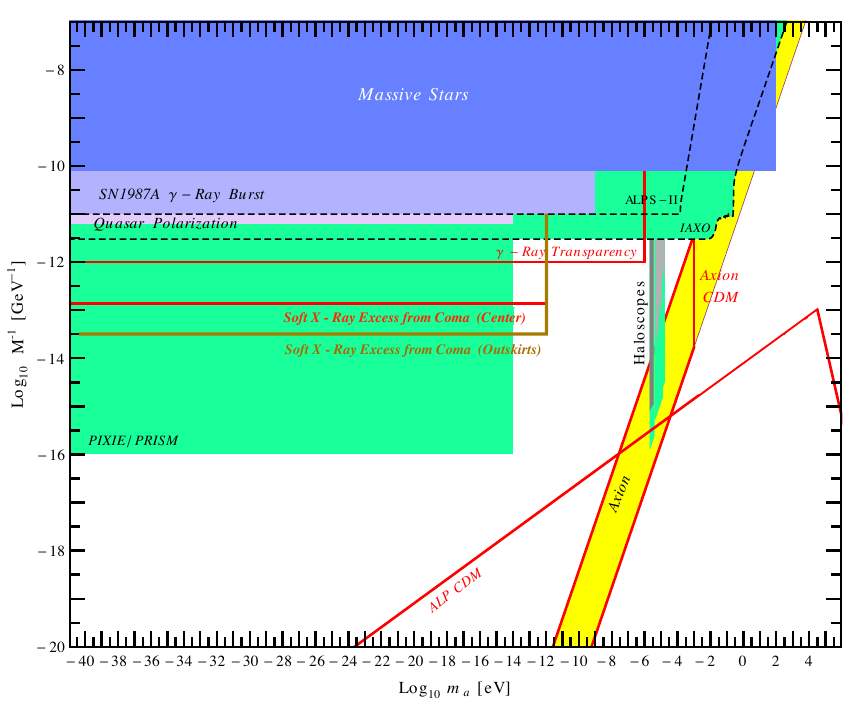}
\caption{The inverse ALP-photon coupling $M^{-1}$ versus its mass $m_a$. The bands for $M^{-1}$ where ALPs with $m_a < 10^{-13}$ eV of a CAB converting to photons can explain the soft X-ray excess of the Coma cluster are shown respectively for the center (red lines) and outskirts (brown lines). We also show exclusions from anomalous energy loss of massive stars~\cite{2013PhRvL.110f1101F}, SN 1987A~\cite{Brockway:1996yr,Grifols:1996id}, a possible bound from quasar polarizations~\cite{Payez:2012vf} and ALPs converting into photons in microwave cavities in magnetic fields~\cite{1987PhRvL..59..839D}. Furthermore, we include the parameters where axions or ALPs can account for all or part of cold dark matter (CDM)~\cite{Arias:2012az} or explain the cosmic $\gamma$-ray transparency~\cite{DeAngelis:2007dy}. The yellow band corresponds to the QCD axion. The green regions mark the parameter space that is expected to be explored by the light-shining-through-wall experiment ALPS-II, the helioscope IAXO, the haloscopes ADMX and ADMX-HF and the CMB experiments PIXIE or PRISM. This figure is extended from~\cite{Dias:2014osa}.}
\label{ALPconstraints_fig}
\end{figure}

As a CAB would be universal, an obvious next step in testing the CAB explanation for the soft excess is
to extend the analysis of~\cite{2013arXiv1312.3947A} to the soft excess in the outer regions of the Coma cluster \cite{Bonamente:2009ns}. It is an essential consistency requirement
to check that the soft excess in the center and the outskirts can be simultaneously explained by the same $M$ and $\langle E_{CAB} \rangle$, at least within the
astrophysical uncertainties set by the magnetic field. As the cluster magnetic field is less constrained in the outskirts than in the center we determine a range of parameters $M$ and $\langle E_{CAB} \rangle$ and show that this range is consistent with the parameters given in~\cite{2013arXiv1312.3947A}. We summarize the range of allowed ALP-photon couplings $M$ together with other constraints on the ALP parameter space in Figure~\ref{ALPconstraints_fig}.

This paper is organised as follows. Section \ref{softex_sec} deals with the determination of the soft excess in the Coma outskirts obtained from the ROSAT mission,
 and the procedure of converting the detector counts into photon flux. Section \ref{axconv_sec} discusses the physics of ALP conversion, and the models used
 for the electron density and magnetic field structure in the Coma outskirts. Section \ref{result_sec} contains results and gives constraints on the ALP-photon coupling and the mean CAB energy required to explain the soft excess in the Coma outskirts.

\section{Existence and Magnitude of the Soft Excess in the Coma Outskirts}\label{softex_sec}

In this section, we want to quantitatively extract the soft excess from the Coma Cluster outskirts by analysing the X-ray data from the ROSAT All-Sky Survey (RASS), following~\cite{2003ApJ...585..722B,Bonamente:2009ns}. The main objective is to calculate the flux/luminosity of observed soft excess in the outskirts of the Coma cluster (in~\cite{2003ApJ...585..722B,Bonamente:2009ns} the soft excess is only given in terms of counts s$^{-1}$ at the PSPC detector).

Results of the RASS are available in terms of position dependent count rates of the ROSAT Position Sensitive Proportional Counter (PSPC) detector. The data is stored in maps with linear pixel size of $12'$ in seven energy bands R1-R7~\cite{Snowden:1997ze}.
 The highest energy band is the R7 band ($1.05-2.04$ keV) and the lowest fully trustable energy band is the R2 band ($0.14-0.284$ keV).\footnote{Although the lowest channels in the
 R1 band suffer from event loss~\cite{1994ApJ...424..714S}, the R1 band does also show a clear soft excess halo around Coma. R3 is not a useful band for analysis.}

In order to calculate the intrinsic flux coming from the Coma cluster the following steps are necessary:
\begin{itemize}
\item \textit{Background subtraction:} The background to emission from Coma itself is either the local or extragalactic diffuse soft X-ray background.
This background is both position-dependent and - due to charge exchange with the solar wind - also time-dependent.
It may be subtracted by considering observations of regions peripheral to the cluster. As the RASS observations come from the satellite slewing across the sky,
this gives a co-temporal and co-spatial background measurement.
More specifically, the local background to the Coma cluster was estimated in~\cite{Bonamente:2009ns} from the $4^{\circ}$ to $6^{\circ}$ annulus with respect to the Coma center. At $4^{\circ}$, the PSPC count rates drop to a value that remains constant up to $6^{\circ}$, while regions further out are affected by emission from the North Polar Spur and hence less suitable for background extraction. Of presently available X-ray data archives, only the RASS allows simultaneous signal and background measurement, as it involved an all-sky
 survey with a large field-of-view instrument (the ROSAT field of view was almost $1^\circ$ in radius).
\item \textit{Fitting the thermal ICM component:} The soft excess is observed in the low energy channels (specifically the R2 band) of the ROSAT PSPC.
More precisely, the count rates in the R7 band ($1.05-2.04$ keV) can be explained by thermal emission of the intra-cluster medium. The PSPC count rates from the R7 band can then
be used to fit the normalisation of the thermal ICM component, in turn allowing the expected thermal component at lower energies in the R2 band to be calculated and compared with the observed PSPC count rates to extract the soft excess.
\item \textit{Fitting the soft excess component:} For distances of up to $\sim 4^\circ$ from the cluster center, the R2 band ($0.14-0.284$ keV), is found to contain significantly higher count rates than predicted from thermal ICM emission \cite{Bonamente:2009ns}, that is, a soft excess was detected. Since the physical origin of the soft excess is not clear, we define different spectral models for the excess corresponding to different possible origins of the soft excess. Using the PSPC count rates in the R2 band, we can then determine the normalisation of the excess component such that the unabsorbed excess flux/luminosity can be calculated for each model of the excess.
\end{itemize}
We can model the overall differential photon flux above background as
\begin{equation}
dN\left(E(1+z)\right) = e^{-n_H \sigma \left(E(1+z)\right)} \left[ C_{\text{ICM}} \frac{g\left(E(1+z),T\right) e^{-\frac{E(1+z)}{k T}}}{\sqrt{kT}\,E(1+z)} + C_{\text{Ex}} X\left(E(1+z)\right)\right] dE\,.\label{genflux2}
\end{equation}
In this equation $z=0.024$ is the redshift of the Coma cluster. As neglecting red-shift dependence in~\eqref{genflux} is only a
 percent level shift that is small compared to other uncertainties, we shall henceforth drop the redshift dependence, giving
\begin{equation}
dN \left(E\right) = e^{-n_H \sigma \left(E \right)} \left[ C_{\text{ICM}} \frac{g\left(E, T\right) e^{-\frac{E}{k T}}}{\sqrt{kT}\,E} + C_{\text{Ex}} X\left( E \right)\right] dE\,.\label{genflux}
\end{equation}
 The exponential factor on the LHS of~\eqref{genflux} models the effective absorption of extragalactic X-rays by neutral hydrogen HI in our galaxy and
 $\sigma\left(E\right)$ is the corresponding photo-electric cross section which is much larger in soft X-rays (R2) than in hard X-rays (R7). $n_H$ is the Milky Way
  neutral hydrogen column density. We use $n_H = 0.9 \cdot 10^{20}\, \text{cm}^{-2}$~\cite{2002ApJ...576..688B} for the entire Coma cluster. Maps of the neutral hydrogen
   distribution have been constructed using 21-cm data, e.g., by Dickey \& Lockman (DL)~\cite{Dickey:1990mf} and more recently by the Leiden/Argentine/Bonn (LAB) group~\cite{Kalberla:2005ts}. For a $5^\circ$ cone-region around the Coma center the HI distribution reported by these two groups, $0.8-1.1 \cdot 10^{20}\, \text{cm}^{-2}$ (DL) and $0.8-1.2 \cdot 10^{20}\, \text{cm}^{-2}$ (LAB), shows no evidence of large gradients and is consistent with the above quoted value of $n_H$.

$g\left(E,T\right)$ is the Gaunt factor, a slowly varying function of $E$. $C_{\text{ICM}} = a\, Z^2 n_{I} n_e$ is the normalisation of the ICM thermal component in~\eqref{genflux} where $a$ is a known numerical factor, $Z$ is the ionic charge, $n_I$ the ion density, and $n_e$ the electron density.
The known dependence of $C_{\text{ICM}}$ on $n_e$ will allow us a crosscheck of our fitting results for $C_{\text{ICM}}$ as $n_e$ is known for the Coma cluster in terms of a $\beta$-model~\cite{1978A&A....70..677C} from ROSAT data~\cite{1992A&A...259L..31B}. $T$ is the ICM temperature. We set it to $T=4$ keV for the regions $0.5-2^{\circ}$ and $T=2$ keV for the regions $2-4^{\circ}$, following the averaged estimates for the plasma temperature given in~\cite{Bonamente:2009ns}.

While \eqref{genflux2} approximates the continuum emission, an accurate fit
 requires inclusion of the line spectra that are present in the plasma gas. The amplitudes of the lines depend on the abundances of the trace elements in particular metals.
We use the fitting program Xspec~\cite{xspec}, where the metal abundance can be specified by a single parameter $A$ that sets the abundance
of metals relative to their solar abundance. Following the best fit values of~\cite{2003ApJ...585..722B}, we use $A=0.2$ in an APEC emission spectrum. Both the used abundance $A$ and the ICM temperatures $T$ are in agreement with a more recent study of the Coma outskirts up to 2 Mpc using Suzaku observations~\cite{2013ApJ...775....4S}.

The excess component in~\eqref{genflux} is parametrized by a normalisation $C_{\text{Ex}}$ and a spectral shape $X\left(E\right)$. We use the following forms for
 $X(E)$ corresponding to different physical origins of the soft excess:
\begin{itemize}
\item{CAB component:}
\begin{equation}
X\left( E \right) = \sqrt{ \frac{E}{E_{CAB}} } \, e^{-\left(\frac{E}{2E_{CAB}}\right)^2}\,,\label{CABspec}
\end{equation}
where $E_{CAB}$ is the characteristic energy of the CAB.
Note that~\eqref{CABspec} is an approximate expression for the ALP energy spectrum. The exact spectral shape can be extracted numerically from the time dependence of the scale factor~\cite{Conlon:2013isa} but does not deviate strongly from~\eqref{CABspec}. However, \eqref{CABspec} is only a good approximation for the \emph{photon} spectrum originating from ALP-photon conversion if the conversion is mostly independent of the ALP energy. This only holds in the so-called small angle regime, which as shown in~\cite{2013arXiv1312.3947A}
does not hold in the central region of the cluster. However we will show in Section~\ref{axconv_sec} that
the small angle approximation is mostly valid in the outskirts of the Coma cluster. In this section, we choose $E_{CAB}=0.188$ keV which corresponds to a mean CAB energy of $0.272$ keV, typical for a CAB originating from late time modulus decays with masses of $\mathcal{O}(10^6)$ GeV, considered in~\cite{Conlon:2013isa}.
\item \textit{Additional warm thermal component:}

In this case, we assume the excess originates from another gas component with a lower warm temperature $T_{\text{soft}}$. Since $T_{\text{soft}}\simeq 0.04-0.1$ keV is much smaller than the temperature of the hot ICM $T=2-4$ keV, this component does not produce significant luminosities at high X-ray energies but only e.g., in the R2 band.
While the continuum emission would have the thermal bremsstrahlung form given in \eqref{genflux}, the spectrum is line dominated and so is modelled via Xspec.
We fit a warm gas with $T_{\text{soft}}=0.08$ keV and $A=0.3$. These values are chosen for comparability with the analysis of~\cite{2002ApJ...576..688B,2013arXiv1312.3947A}.

We are ultimately interested in trying to explain simultaneously the soft excess in the center and the outskirts of Coma through ALP-photon conversion.
The phenomenological constraints in~\cite{2013arXiv1312.3947A}, for example on the ALP-photon coupling, were
based on the soft excess luminosities extracted in~\cite{2002ApJ...576..688B}. These were obtained by modeling the soft excess as originating from a warm gas with the above specified $T_{\text{soft}}$ and $A$. Hence, for a consistent comparison of excess luminosities in the outskirts of Coma to those in the center,
we have to extract the excess luminosity in the outskirts in the same manner: that is, by formally treating the excess as originating from a warm gas.
For this reason, the extracted warm gas soft excess luminosities will be used in Section~\ref{result_sec}.
Finally, note that $A=0.3$ does not correspond to the most conservative fit in terms of luminosities for the soft excess which would be $A=0$. The luminosities for these two different values of $A$ are expected to differ by $\mathcal{O}(10\%)$~\cite{Bonamente:2009ns}.
\item \textit{Power law component:}
\begin{equation}
\label{adg}
X\left(E\right) = E^{-\alpha}\,,
\end{equation}
with photon index $\alpha$, as would apply for an origin of the soft excess via inverse Compton scattering.
For definiteness, we take the index $\alpha$ to be
 $\alpha = 1.75$~\cite{1999ApJ...517L..91L}. The main phenomenological difficulty with the inverse Compton scattering scenario is that relativistic electrons
 also necessarily emit in radio via synchrotron emission and in gamma-rays through non-thermal bremsstrahlung. Extrapolation of \eqref{adg} to higher energies implies over-production
 of radio emission, necessitating a very sharp cutoff in the spectrum, and even then there is a problem with a failure to observe clusters in gamma rays.
  For a detailed discussion of the difficulties to explain the soft excess phenomenon from known astrophysical processes in clusters see~\cite{2013arXiv1312.3947A}.
\end{itemize}

\begin{figure}[t!]
\centering
\includegraphics[width=0.49 \linewidth]{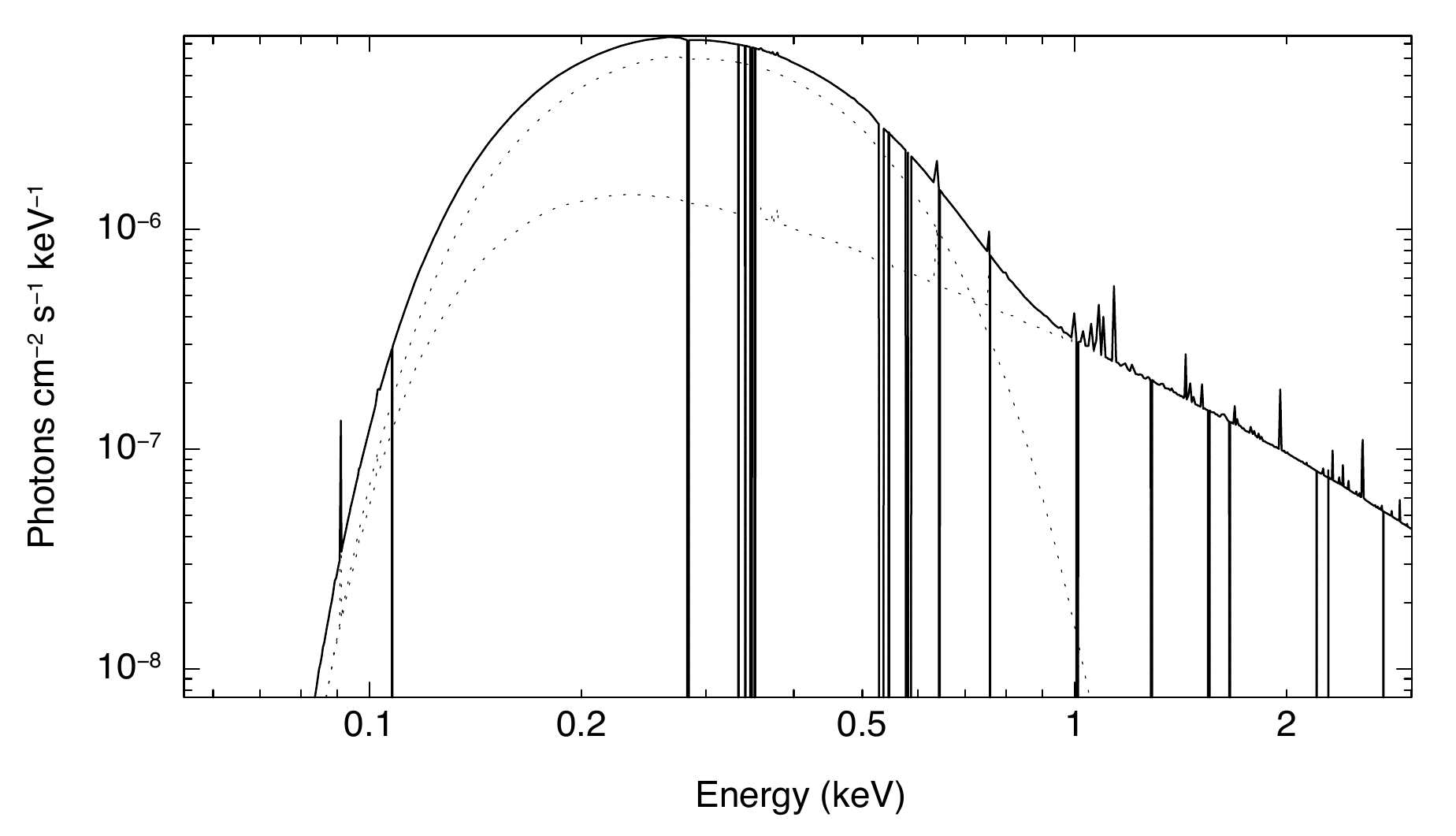}
\includegraphics[width=0.49 \linewidth]{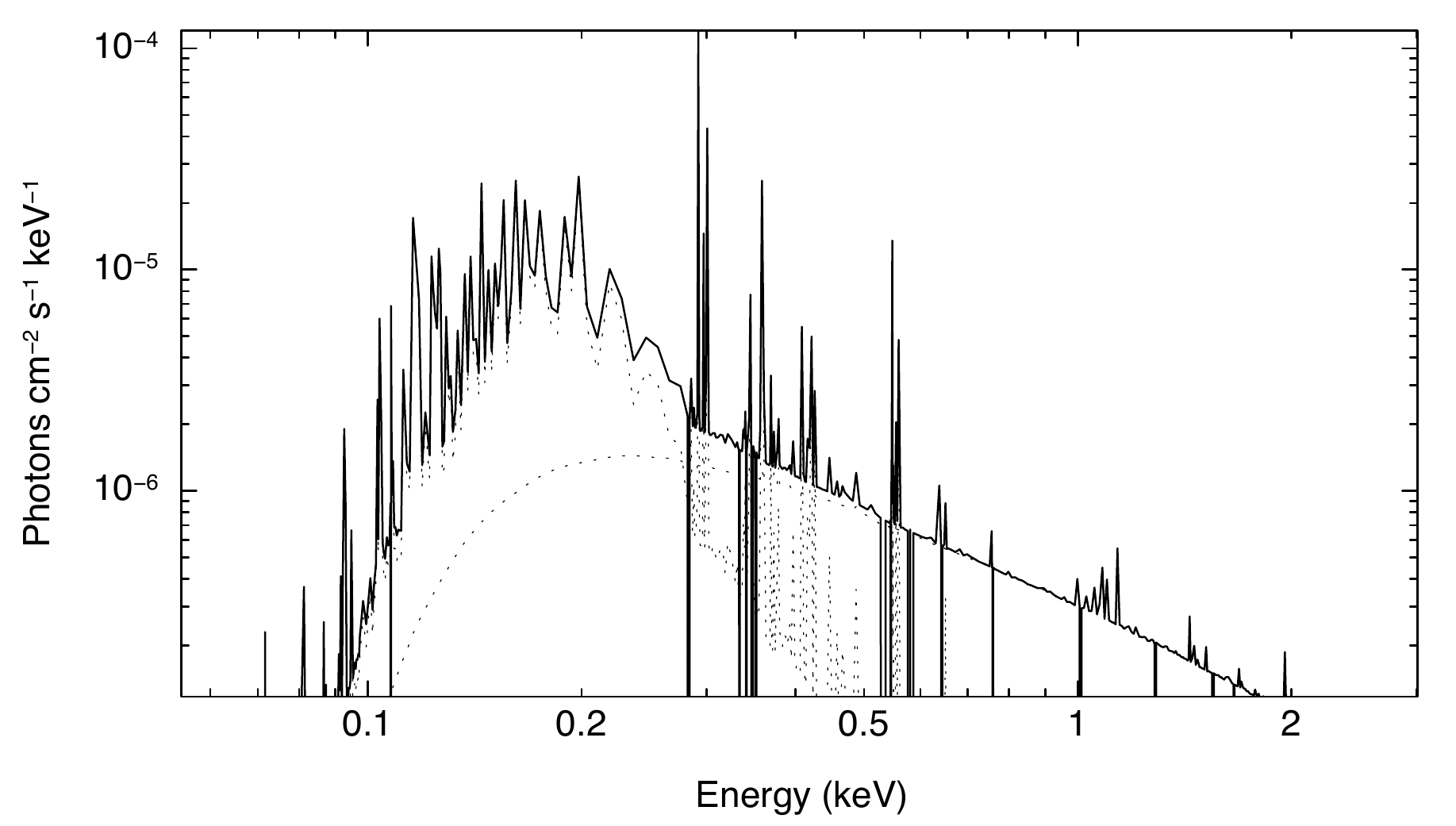}
\caption{Differential photon flux for the $1^\circ-1.5^{\circ}$ region. The shown spectra are CAB (left) and thermal (right).}
\label{fits_fig}
\end{figure}

\begin{figure}[h!]
\centering
\includegraphics[width=\linewidth]{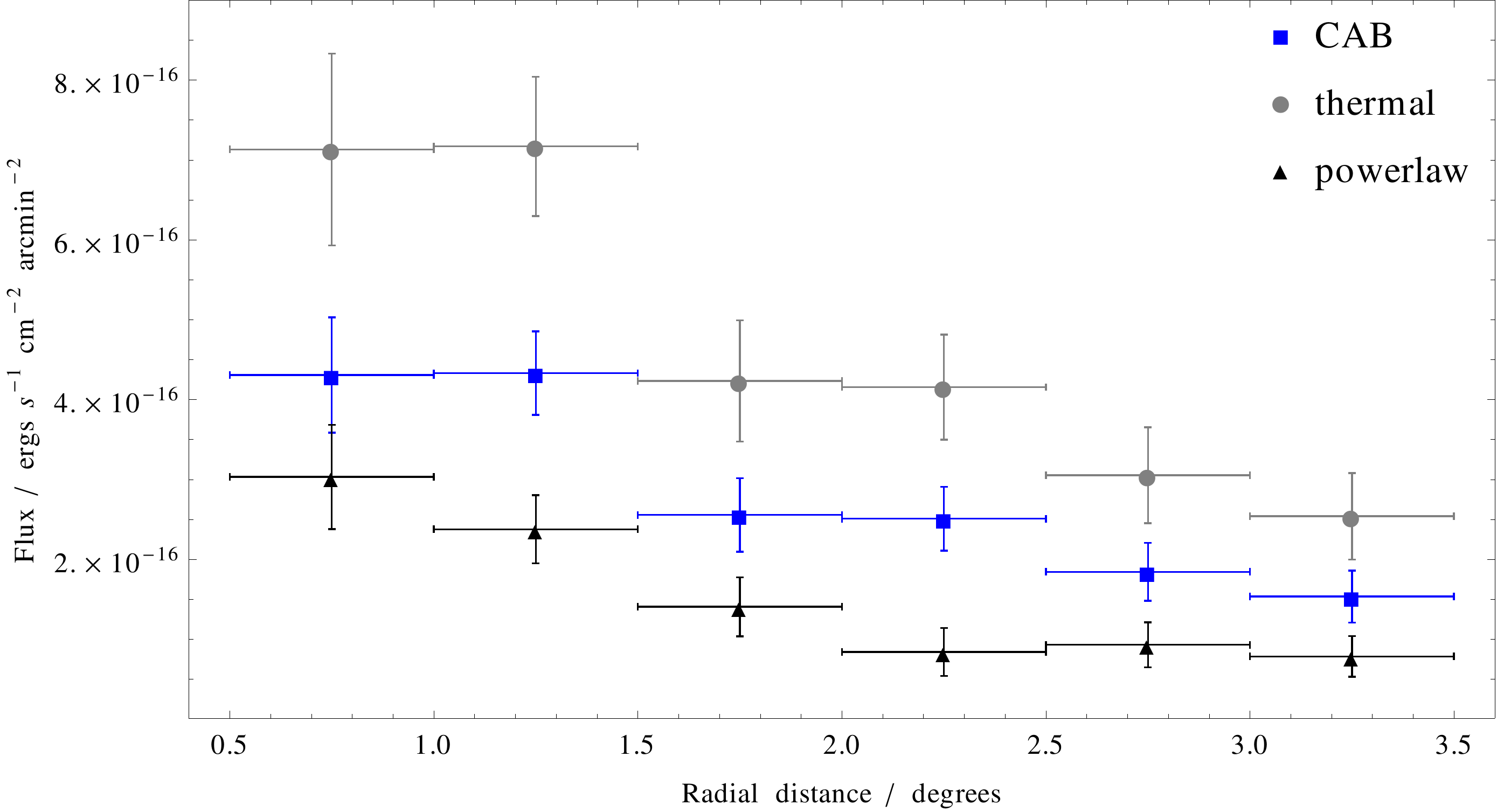}
\caption{Measured flux of the excess in the R2 band ($0.14-0.284$ keV) from the RASS
depending on the different models for the spectral shape of the excess.}
\label{excessmorph_fig}
\end{figure}

We give exemplary energy spectra for two of the three different spectral shapes in Figure~\ref{fits_fig}, for which the excess component has either the CAB form or a thermal form. The spectral lines of the thermal components are clearly visible as local peaks above the continuous emission spectrum.
The fitting procedure can be repeated for the different regions of the Coma cluster outskirts, up to $3.5^{\circ}$, allowing us to extract the
morphology of the excess in the R2 band, depending on the different models for the excess component, as shown in Figure~\ref{excessmorph_fig}.
At even larger distances the excess disappears and merges into the background~\cite{Bonamente:2009ns}.
The errors on the flux in Figure~\ref{excessmorph_fig} correspond to the 1-$\sigma$ fitting errors returned by Xspec.
Figure~\ref{excessmorph_fig} shows that
the excess luminosity obtained has a strong dependence on the assumed physical origin of the soft excess. This is in agreement with~\cite{2002ApJ...576..688B}, where it was found that assuming either a thermal (warm gas) or a power-law (inverse Compton) soft excess origin, the obtained excess luminosities typically differ by a factors of $\mathcal{O}(2-3)$, and up to an order of magnitude in extreme cases.

Under the assumption that the soft excess is explained by a photon spectrum~\eqref{CABspec} originating from a CAB, the fitting procedure can be used to bound $E_{CAB}$ - or equivalently the mean CAB energy - from above. Raising $E_{CAB}$ corresponds to shifting the CAB peak in Figure~\ref{fits_fig} to higher energies, and
 above a certain $E_{CAB}^{\text{max}}$ there will be significant energy deposition in the R7 band. This is undesirable since the R7 emission can be solely explained by thermal ICM emission. We find that the quality of the overall fit to the R2 and R7 spectrum worsens significantly for $\langle E_{CAB} \rangle > \langle E_{CAB} \rangle^{\text{max}}  \simeq 0.37$ keV in all five regions that the cluster outskirts have been divided into.

\section{Predicted Excess from ALP conversion}\label{axconv_sec}

Our aim is to see whether the excess soft
X-ray halo around the Coma cluster 
can be explained by the conversion of ALPs into photons.
ALPs convert to photons in homogeneous magnetic fields, with a mixing that is set by the difference between the ALP mass and the effective photon mass
(the plasma frequency). The computation of ALP-photon mixing therefore requires knowledge of both the magnetic field and the electron density.
We first describe our model for the electron density in the Coma outskirts (Section~\ref{ICM_subsec}) and then describe our model for the magnetic field
(Section ~\ref{Bfield_subsec}). In Section~\ref{Consistency_sec} we perform some consistency checks to show that our numbers are reasonable, before finally describing in Section
\ref{axioprob_subsec} how we compute the probability of ALP to photon conversion for a given astrophysical model.

\subsection{Density profile of hot gas in Coma}\label{ICM_subsec}

The Coma cluster has a complex structure when examined in detail \cite{2003A&A...400..811N}. However,
 the broad X-ray picture of the cluster is simpler. It consists of a roughly spherical central region, with the
 merging NGC4839 group located about $0.6^\circ$ south-west from the centre and some emission in between (e.g., see Figure~1 of \cite{1992A&A...259L..31B}).
 This suggests the use of a simple analytical model to describe the cluster, consisting of the sum of two $\beta$-models.

X-rays emitted from clusters come chiefly from the intracluster medium (ICM), a hot plasma, via thermal bremsstrahlung.
Good fits for the electron density are obtained from the $\beta$-model \cite{1978A&A....70..677C}:
\begin{equation}
n_e(r)=n_0\left( 1+ \frac{r^2}{r_c^2} \right)^{-\frac{3}{2}\beta}.
\end{equation}
The expression is inspired by considering an isothermal cluster in hydrostatic equilibrium. The parameters $r_c$ and $\beta$ of the $\beta$-model are empirical, allowing for the accurate determination of the gas density even when the isothermal-hydrostatic assumption is not valid \citep{1999ApJ...511...65J}.

Using ROSAT to fit the surface brightness, best fit parameters were found by \cite{1992A&A...259L..31B} to be $\beta=0.75 \pm 0.03$, $r_c=291 \pm 17$\,kpc and $n_0=3.44 \pm 0.04 \cdot 10^{-3} \, \text{cm}^{-3}$.
This fit was performed up to a distance of about 100\,arcmin ($1.67^\circ$ or $2.8 \, \text{Mpc}$ from the centre). The central density $n_0$ is a derived quantity from the best-fit central surface brightness \citep{1992A&A...259L..31B}.

Another study of the Coma X-ray surface brightness (with XMM-Newton) \cite{2012MNRAS.421.1123C} focused on the core region (central 1000\,arcsec $\sim 0.3^\circ$ ) of the cluster. They found the parameters for the $\beta$-model to be $\beta = 0.6$ and $r_c=245 \, \text{kpc}$. Within the central region the ROSAT and XMM-Newton fits,
assuming the same central density, are consistent with each other (less than 5\% difference). An older fit to the Coma cluster using the Einstein Observatory within the central 0.2 degrees \cite{1999ApJ...511...65J}, once corrected for cosmology, results in $\beta=0.67$ and $r_c=0.31$\,Mpc. This density profile is again broadly consistent with the ROSAT and XMM-Newton studies and the more recent Suzaku observations of the Coma cluster~\cite{2013ApJ...775....4S}.

To model the electron density in the outskirts up to distances of around 4 degrees or $6.8$\,Mpc, we use the $\beta$-model evaluated at these radii.
This gives an estimate for the electron density there as $n_e(6 \, \text{Mpc}) \sim 6 \cdot 10^{-6} \,  \text{cm}^{-3}$.
This region is part of the Coma supercluster, and these electron densities are typical of those expected from supercluster regions, and is an
order of magnitude above the mean density of hydrogen nuclei in the universe  $\bar{n}_H=\Omega_b \frac{\rho_{crit}}{m_H}(1-Y)(1+z)^3 \approx 2 \cdot 10^{-7} \, \text{cm}^{-3}$.\footnote{$\Omega_b$ is the baryon fraction in the universe, $\rho_{crit}$ the critical density of the universe and $Y$ is the Helium abundance.}
This suggests that the model for the electron density is meaningful at such large radii and does not produce results which are physically implausible.

As the $\beta$-model is extended beyond the infalling NGC4839 group, the contribution of this group to $n_e$ and consequently to the magnetic field needs to be included. Little is known about the plasma distribution in the group. The mass of NGC4839 is $\sim 0.1$ of the Coma cluster \cite{2001A&A...365L..74N}. It was modeled by \cite{2013MNRAS.433.3208B} as another $\beta$-model localised at the position of NGC4839 scaled in a self-similar way from the model for the central part of the cluster with NGC4839 $\beta$-model parameters of $n_0=3.44 \times 10^{-3}\, \text{cm}^{-3}$, $\beta=0.75$ and $r_c=134  \, \text{kpc}$. Away from the group the double-$\beta$ model (Coma+NGC4839) quickly converges to the single-$\beta$ model fitted by excluding the group. It also agrees well with the gas density profile obtained by Suzaku observations in the direction of NGC4839 (see Figure 14 in \cite{2013MNRAS.433.3208B}).
For this paper we use the double-$\beta$ model.

\subsection{Magnetic field model in the outskirts of Coma}\label{Bfield_subsec}

As we discuss below in Section \ref{axionconv}, the magnitude of ALP-photon conversion depends on the square of the magnetic field.
The first evidence for the magnetic field in the Coma cluster came from the diffuse radio halo \cite{1970MNRAS.151....1W} associated with
synchrotron radiation that extends beyond the central 1\,Mpc of the cluster. The magnitude of synchrotron emission is degenerate between the
density of the relativistic electron population and the strength of the magnetic field. The equipartition assumption can be used to break this degeneracy,
  leading to an estimate of $B \sim 0.7-1.9 \, \mu$G \cite{2003A&A...397...53T}, averaged over the central 1\,Mpc$^3$. A potential observational method to break the degeneracy  is by directly observing the relativistic electron population via a hard X-ray signal from inverse Compton scattering of CMB photons off the relativistic electrons.
   The lack of such non-thermal hard X-ray emission from Coma then places a lower bound on the average magnetic field of $B > 0.2 \, \mu {\rm G}$ \cite{2004ApJ...602L..73F,2008MNRAS.383.1259C}.

A different method for determining the magnetic field comes from Faraday rotation of linearly polarised light. The ICM plasma and the magnetic field induce different phase velocities for right-handed and left-handed circularly polarised light. This causes a wavelength dependent rotation of the plane of polarisation for linearly polarised light coming from localised radio sources.
\begin{equation}
\Psi_{obs}(\lambda) = \Psi_0 + \lambda^2 ~ RM,
\end{equation}
where  $\Psi$ is the angle of polarisation, $\lambda$ the frequency of light and
\begin{equation}
RM = \frac{e^3}{2\pi m^2_e} \int_{l.o.s} n_e(l)B_{\parallel}(l) \mathrm{d} l\,,
\end{equation}
is the rotation measure. The Faraday rotation method probes the component of the magnetic field parallel to the line of sight multiplied by the electron density. To constrain the magnitude, simulated magnetic fields with a given spectrum are used to produce mock RM images which are then compared with the measured ones~\cite{2004A&A...424..429M}. This in turn provides the perpendicular component of the field which is relevant for ALP conversion (Section~\ref{axionconv}). 

Radio halo observations and magneto-hydrodynamics simulations suggest the magnitude of the magnetic field is attenuated with distance from the cluster centre \cite{2004A&A...424..429M}. Therefore the radial dependence of the absolute value of the magnetic field is modelled as a scaling of the electron density,
\begin{equation}
B(r)=C \cdot B_0\left( \frac{n_e(r)}{n_0} \right)^{\eta}\,,
\end{equation}
where the constant $C$ is chosen such that  $B_0$ corresponds to the average magnetic field in the core of the cluster. The $\eta$ parameter is determined empirically (e.g., through fitting Faraday rotation measures \cite{2010A&A...513A..30B,2004A&A...424..429M,2013MNRAS.433.3208B}). Theoretically motivated values come from either the isothermal result, $B(\mathbf r) \propto n_e(\mathbf r)^{\frac{1}{2}}$ or the case where the magnetic field is `frozen' into matter $B(\mathbf r) \propto n_e(\mathbf r)^{\frac{2}{3}}$.

The actual magnetic field is turbulent and multi-scale. It can be modelled as a Gaussian field with a power spectrum
$\langle \vert \tilde{B}(k) \vert ^2 \rangle \propto \vert k \vert^{-n+2}$ over a range of scales between $k_{min} = 2\pi / \Lambda_{max}$ and $k_{max} = 2\pi / \Lambda_{min}$. The
 magnetic field then has structure between the two scales $\Lambda_{max}$ and $\Lambda_{min}$.

In \cite{2010A&A...513A..30B}, Faraday rotations measures within 1.5 Mpc from the Coma cluster centre were used to constrain models of the magnetic field.
The best fit values for the central magnetic field and the $\eta$ parameter were
$B_0=4.7 \, \mu$G and  $\eta = 0.5$, with a $1\sigma$ range between ($B_0=3.9 \,\mu$G; $\eta = 0.4$) and ($B_0=5.4 \, \mu$G; $\eta = 0.7$).
There is a degeneracy between the power-law index $n$ and the maximum coherence scale $\Lambda_{max}$. The data can be fitted by
 a Kolmogorov spectrum ($n=17/3$) with scales between $\Lambda_{min}=2 \, \text{kpc}$ and $\Lambda_{max}=34 \, \text{kpc}$, but equally well by a flat spectrum ($n=4$)
 with coherence lengths between $\Lambda_{min}=2 \, \text{kpc}$ and $\Lambda_{max}=100 \, \text{kpc}$, and
 ($B_0=5.4 \, \mu$G; $\eta = 0.7$). These two models are summarised in Table~\ref{table:models}.

Our description of the magnetic field will be based on these models, with the radial parameter taken to the outskirts region.
On general grounds, the coherence length is expected to grow as one moves to the outskirts and the electron density decreases.
We will analyse this by considering two extreme cases.
For equilibrium cool-core clusters the characteristic turbulence length scale has been argued to grow as $L \propto n_e^{-1}$~\cite{2011MNRAS.410.2446K}. The best fit for the magnetic field profile from Faraday rotation measures coincides with the isothermal scaling ($\eta = 0.5$) and the spectrum that well describes the data is Kolmogorov. Hence this scaling of the characteristic length ($L \propto n_e^{-1}$) is adopted as an extremal case that could apply to the Coma cluster. The other case is where the coherence lengths stay the same all the way to the outskirts of the cluster, with the most adequate description being somewhere between the two extremes. In the case where the characteristic length scale grows with radius its value is fixed by specifying the average coherence lengths within the cluster core.

\begin{table}
\begin{center}
\begin{tabular}{| l | c | r |}
  \hline
   &Model A& Model B \\
  \hline
  $\Lambda_{min}$ & 2 kpc & 2 kpc \\
  \hline
  $\Lambda_{max}$ & 34 kpc& 100 kpc \\
  \hline
  $n$ & 17/3 & 4 \\
  \hline
  $B_0$ & 4.7 $\mu$G & 5.4 $\mu$G\\
  \hline
   $\eta$ & 0.5 & 0.7 \\
  \hline
\end{tabular}
\caption{Magnetic field models giving good fits for the Faraday rotation measures to the central regions of the Coma cluster. The magnetic field spectrum ranges in wave number from
$\frac{2 \pi}{\Lambda_{min}}$ to $\frac{2 \pi}{\Lambda_{max}}$.}
\label{table:models}
\end{center}
\end{table}

\subsection{Consistency checks in outskirts} \label{Consistency_sec}

We find typical magnetic fields in the outskirts region at about 4 Mpc from the Coma centre to be $B \sim 0.35 \, \mu {\rm G}$ for Model A and $B \sim 0.15 \, \mu {\rm G}$ for Model B.

Let us check that these values are reasonable. There have been a limited number of observational studies of magnetic fields in the outskirts of clusters/ on supercluster scales.
A value of $B \sim 0.5 \, \mu {\rm G}$ was found by \cite{Kim1989} in the study of the bridge region of the Coma cluster, at a distance of around 1.5\,Mpc from the Coma centre. \cite{2007ApJ...659..267K} also finds evidence for $B \sim 0.2-0.4 \,\mu {\rm G}$ magnetic fields on distance scales of 4\,Mpc from the centre of Coma. From the study of Faraday rotation measures in the Hercules-Pisces supercluster, \cite{XuKronberg2006} estimates a magnetic field
of $B \sim 0.3 \, \mu {\rm G}$ and considers typical electron densities in this region as between $5 \ti 10^{-6} \, {\rm cm}^{-3}$ and $2 \ti 10^{-5} \,{\rm cm}^{-3}$.~\cite{Bagchi2002} estimate a magnetic field $B \sim 0.5 - 1 \, \mu {\rm G}$ for a filamentary region of galaxies over a scale $d \sim 6\, {\rm Mpc}$.

Normalization of the thermal component fitted in the R7 band depends on the ICM density, $C_{\text{ICM}} = a\, Z^2 n_{I} n_e$ (see Section~\ref{softex_sec}). Here $a$ is a known numerical factor, $Z$ is the ionic charge, $n_I$ the ion density, and $n_e$ the electron density. The normalization therefore provides a cross-check on the double-$\beta$ model for the electron density we use to describe the Coma cluster. The modelled ICM density is to within a factor of two compared to the density deduced by fitting the thermal component to observations.\footnote{With the exception of $2^\circ - 2.5^\circ$ radial bin where there is no hard thermal component after background subtraction.}

A recent observation of the Coma cluster by Suzaku~\cite{2013ApJ...775....4S} measured the temperature, metallicity and electron density radial profiles along five different directions. The values for $kT$ and metallicity $A$ for the hard thermal component used in this paper are consistent with the observation. Our modelled electron density agrees well with the observed radial profile. The measurement of the X-ray surface brightness along the five directions converges to the same value above 80\,arcmin. Below that the most prominent feature is a bump in the SW direction roughly centred on the NGC4839. Thus the double-$\beta$ model we use also encapsulates the approximate morphology of the Coma cluster.

Based on the above observations, it appears that the electron densities and magnetic field strengths we are using in the outskirts region are
physically sensible. Of course one should probably not trust the magnetic field strengths to within a factor of two, but there does not seem to be any reason to suppose
an order of magnitude error in the values.

\subsection{Axion conversion}\label{axioprob_subsec}
\label{axionconv}
The part of the ALP-photon Lagrangian responsible for the conversion is
\begin{eqnarray}
{\cal L} &\supset &  \frac{1}{M}  a\, \boldsymbol{E} \cdot \boldsymbol{B} \, ,
\end{eqnarray}
where  $M^{-1}$ is the ALP-photon coupling.

The ALP to photon conversion probability for a single domain of homogeneous magnetic field of size $L$ is \cite{Raffelt:1987im}:
\begin{equation}
\label{eq:convprob}
	P(a\rightarrow\gamma)= \sin^2(2\theta)\sin^2\left(\frac{\Delta}{\cos 2 \theta}\right),
\end{equation}
 where $\tan 2 \theta = \frac{2B_{\perp} E}{M m_{\rm eff}^2}$, $\Delta=\frac{m^2_{\rm eff}L}{4E}$, $m_{\rm eff}^2=m_a^2 -\omega_{\rm pl}^2$ and $E$ is the ALP energy.
 We assume that the mass $m_a$ is much smaller than the plasma frequency $\omega_{pl} \sim 10^{-13}$ eV and set it to zero.\footnote{Note that for the ALP considered in this paper the mass $m_a$ and the coupling $M^{-1}$ are not related.}
Numerically these parameters evaluate to
\bea
	\theta & \approx & \frac{B_{\perp}E}{M m_{\rm eff}^2} = 8.1\ti 10^{-5}\left(\frac{n_0}{n_e}\right)\left(\frac{B_{\perp}}{1 \mu G}\right)\left(\frac{E}{200\hbox{\,eV}}\right)\left(\frac{10^{13}\hbox{\,GeV}}{M}\right), \\
\Delta & = & 0.93\left(\frac{n_e}{n_0}\right)\left(\frac{200\hbox{\,eV}}{E}\right)\left(\frac{L}{1 {\rm kpc}}\right),
\eea
where $n_0 = 3.44 \ti 10^{-3}\, {\rm cm}^{-3}$.
 The plasma frequency of the ICM depends on the electron density $\omega_{pl}=\sqrt{\frac{4 \pi \alpha n_e}{m_e}}$.

In the simple and illustrative `small angle approximation' ($\Delta \ll 1$ and $\theta \ll 1$) the conversion probability takes the simple form
\begin{equation}
\label{eq:smallangle}
P(a\rightarrow\gamma) = 2.3\times 10^{-8} \left( \frac{B_{\perp}}{1\mu G} \frac{L}{1\mathrm{kpc}} \frac{10^{13}\hbox{\,GeV}}{M} \right)^2 \, .
\end{equation}
In this approximation the explicit dependence on the photon energy and the electron density disappears. However, the dependence on the electron density remains implicit via the scaling of the magnetic field but the knowledge of the central density $n_0$ is not needed. The spectral shape of the cosmic axion background is preserved in this approximation as the conversion probability is independent of the energy of the ALPs. Note the quadratic dependence of the conversion probability on the relevant quantities ($B,L,M$). $\Delta$ falls with the plasma density whereas $\theta$ is always small for the characteristic values of $B, M$ and $n_e$. Therefore the approximation becomes more valid in the outer regions of the cluster (see Figure \ref{simvssdM1}). For the values of the parameters considered here only the first radial bin ($0.5^{\circ}-1^{\circ}$) falls out of the small angle approximation. However, if the
$L \propto n_e^{-1}$ scaling of the coherence lengths is allowed then the small angle approximation is less valid, and ceases to be a valid approximation which is why we implement a semi-analytical approach discussed below.

\subsubsection*{Validity of the single-domain formula}
The propagation and conversion of ALPs in multiscale magnetic fields is computationally expensive. The outer regions of the Coma cluster have around a thousand times more
volume than the central $1 \, \text{Mpc}^3$ region for which a full numerical simulation of the magnetic field has been performed \cite{2013arXiv1312.3947A}.
As this took a number of days to generate the field and determine the ALP-to-photon conversion probabilities, it is not feasible
to simulate the outer regions at the same detail.
This necessitates the use of semi-analytical estimates for the ALP-to-photon conversion probabilities, based on the use of the single domain formula.

To capture multiscale properties of the magnetic field within the single domain formula, we convolve the conversion probability per unit length
with the distribution for coherence lengths $p(L)$ obtained from the magnetic field spectrum. For a given $L$ the field has a full sinusoidal oscillation. Thus the appropriate range for coherence lengths in the single-domain framework is between $\Lambda_{min}/2$ and $\Lambda_{max}/2$.

The random direction of the magnetic field is modelled by replacing $B_{\perp}$ with $B/\sqrt{2}$ ($B$ being the magnitude of the field).
This replacement is suggested by averaging the small angle single-domain formula (Eq.~\ref{eq:smallangle}) over all directions, and the fact that $\theta$ is always in the
small angle approximation.
For the case that the characteristic coherence length scales inversely with electron density, the distribution $p(L)$ and $\Lambda_{min},\Lambda_{max}$ become
dependent on the position $\mathbf{x}$ in the cluster.

\begin{figure}[h!]
\centering
\includegraphics[width=\linewidth]{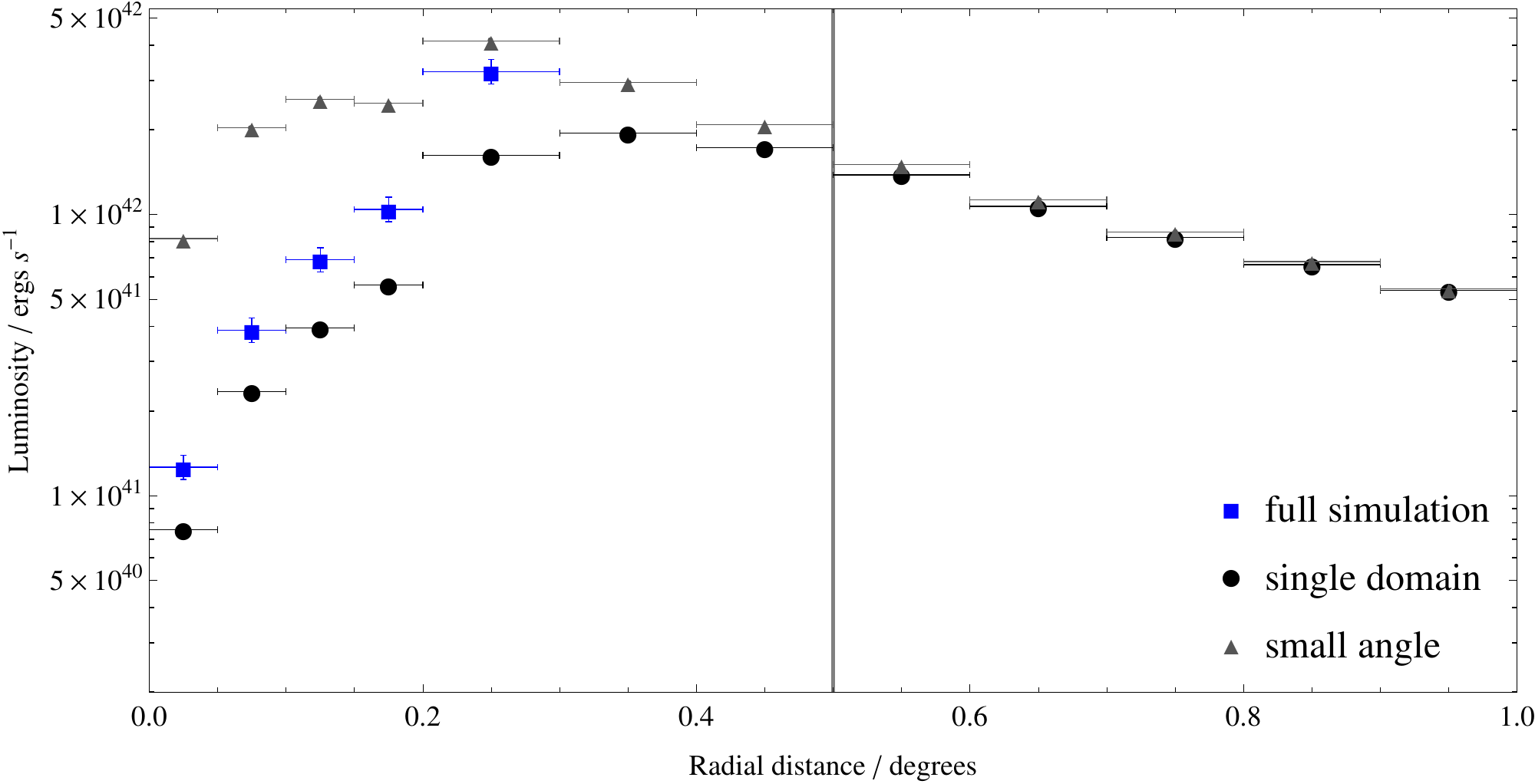}
\caption{Luminosity from ALP-photon conversion for the single domain formula, the small angle single domain formula and the full simulation in the central regions of the cluster for Model A for $M= 6.5 \cdot 10^{12}$\,GeV and $\langle E_{CAB }\rangle = 0.15$\,keV. The vertical line is at 0.5$^{\circ}$, where our analysis of the outer regions begins. The agreement between small angle and single domain approximation in the outskirts improves for larger $\langle E_{CAB }\rangle$ and is still satisfactory for $\langle E_{CAB }\rangle$ at the lower end of the considered values $\langle E_{CAB }\rangle \gtrsim 0.05$\,keV. The errors for the simulation are estimated from the variation of the values when repeating the runs.}
\label{simvssdM1}
\end{figure}

To get a sense of its reliability,
the single domain formula has been tested against the full simulation of the ALP conversion in the
central 1\,Mpc$^3$. Using the single domain formula qualitatively reproduces the radial dependence of the conversion probability and luminosity for the two models considered in this paper (see Figure \ref{simvssdM1}). However, compared to the full simulation, the value of $M$ required to give the same overall magnitude of luminosity differs
 by about 50\% compared to the full simulation in the central $\text{Mpc}^3$ of Coma for both Model A and Model B.
 Given that there are in any case significant astrophysical uncertainties on the magnetic field,
 the single domain formula serves as a reasonable semi-analytic estimate for ALP-photon conversion in the outer regions of the cluster.

The energy density of the CAB is determined via~\cite{Conlon:2013isa}
\begin{equation}
\rho_{CAB} = \Delta N_{eff}\frac{7}{8}\left( \frac{4}{11} \right)^{4/3}\rho_{CMB}\,,
\end{equation}
if the ALPs are the only additional relativistic species in the universe.
The combination of the overall energy density of the CAB, the spectral shape $X(E)$ and the conversion probability
allows the calculation of the spectrum and luminosity of the converted photons.
The prediction for the luminosity from each annular region of the cluster of volume $V$ is:
\begin{equation}
{\mc L} = \int_{V} \int_{\Lambda_{min}(\mathbf{x})/2}^{\Lambda_{max}(\mathbf{x})/2} \int_{E_{min}}^{E_{max}} \frac{c}{L} P(a\rightarrow\gamma ; L, E, \mathbf{x})~ p(L,\mathbf{x}) ~ C_{CAB} ~ E ~ X_{CAB}(E) ~ \mathrm{d}E ~ \mathrm{d}L ~ \mathrm{d}\mathbf{x}^3\,,
\end{equation}
where $C_{CAB}$ is a normalization constant such that $\rho_{CAB} = C_{CAB} \int dE\,E\,X_{CAB}(E)$.
This predicted luminosity (and the associated flux) is the quantity compared with the flux obtained by spectral fitting to ROSAT data.

\section{Results}\label{result_sec}
Once the astrophysical model for the outer regions of the Coma cluster is fixed, with the radial dependence of the magnetic field and its spectrum
 specified, the measured soft excess can be used to constrain the ALP-photon coupling $M^{-1}$ (we recall that
 we assume the ALP mass to be much smaller than the plasma frequency, $m_a \ll \omega_{pl} \sim 10^{-13}$\,eV.)

For each of the two models for the magnetic field we consider two possibilities for the coherence lengths as discussed above.
 We either scale the typical coherence length inversely with the electron density, or retain the same range of coherence lengths
 as in the core of the cluster.
This leads to slightly different predicted luminosities for each annular bin.
For each $\left\langle E_{CAB} \right\rangle $ the predicted fluxes from the ALP conversion are calculated. For a given value of $\left\langle E_{CAB} \right\rangle $,
 the spectrum is used to extract the predicted flux for ROSAT. Demanding that the predicted fluxes lie within the one sigma error\footnote{Using two sigma errors instead results in at most 10\% change in the values of $M$.} of the observed fluxes determines the range for $M$ for which that is true. This gives allowed ranges for $M$ that can explain the morphology and the magnitude of the soft excess as a function of the impact parameter (i.e., annular bins). More conservatively, the range for $M$ gives the approximate lower bound on $M$, as any value above that will not overproduce soft X-rays.

\begin{figure}[h!]
\centering
\includegraphics[width=\linewidth]{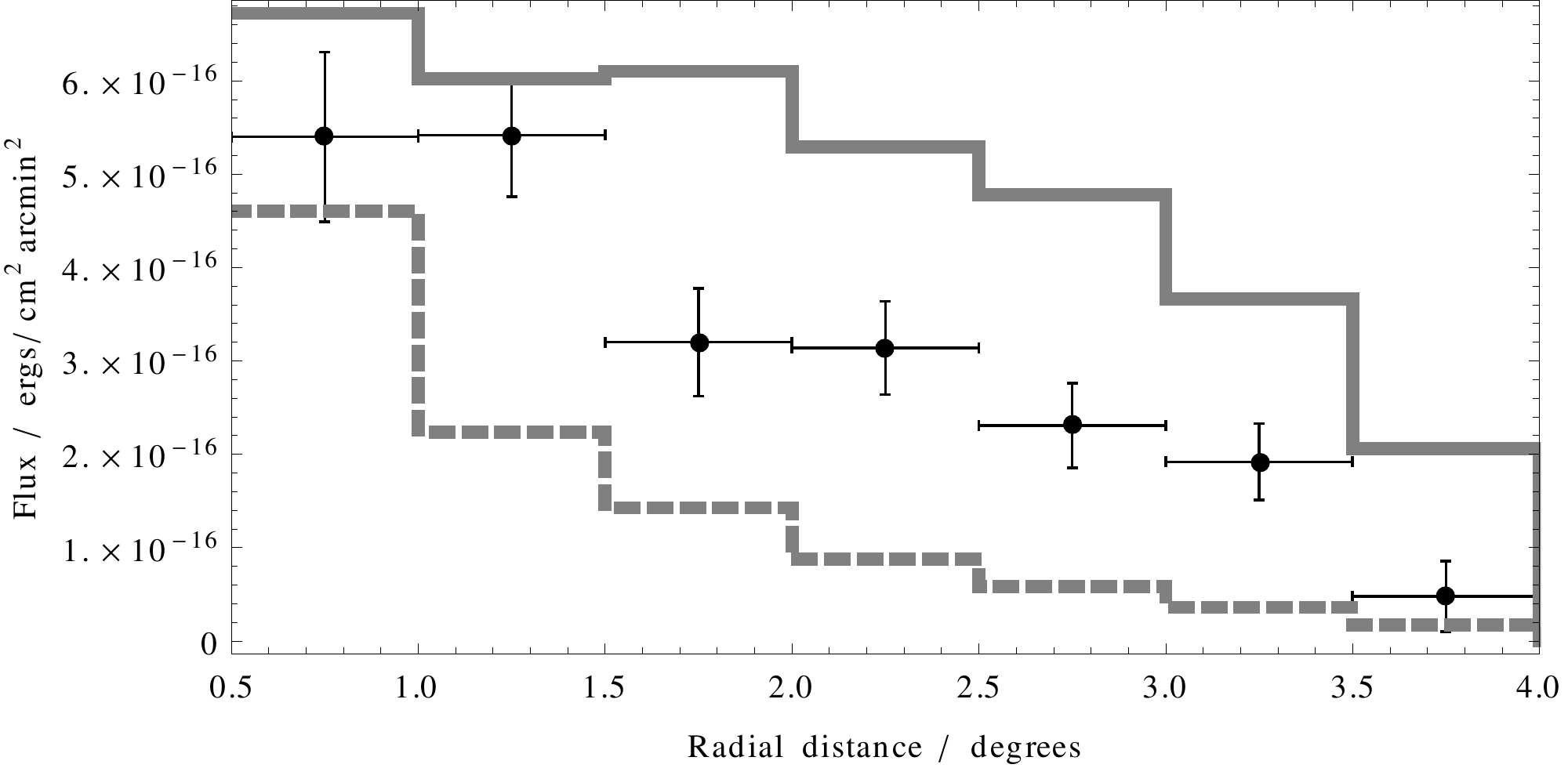}
\caption{Predicted fluxes for Model A magnetic field with $\left\langle E_{CAB} \right\rangle = 0.18$\,keV. Solid line corresponds to $M=3.1\times10^{13}$\,GeV with the scaling of the coherence lengths $L\propto n_e^{-1}$. Dash line is $M=2.0\times10^{13}$\,GeV with the coherence lengths the same throughout the cluster. Note how the measured fluxes lie between the two cases.}
\label{model1}
\end{figure}

\begin{figure}[h!]
\centering
\includegraphics[width=\linewidth]{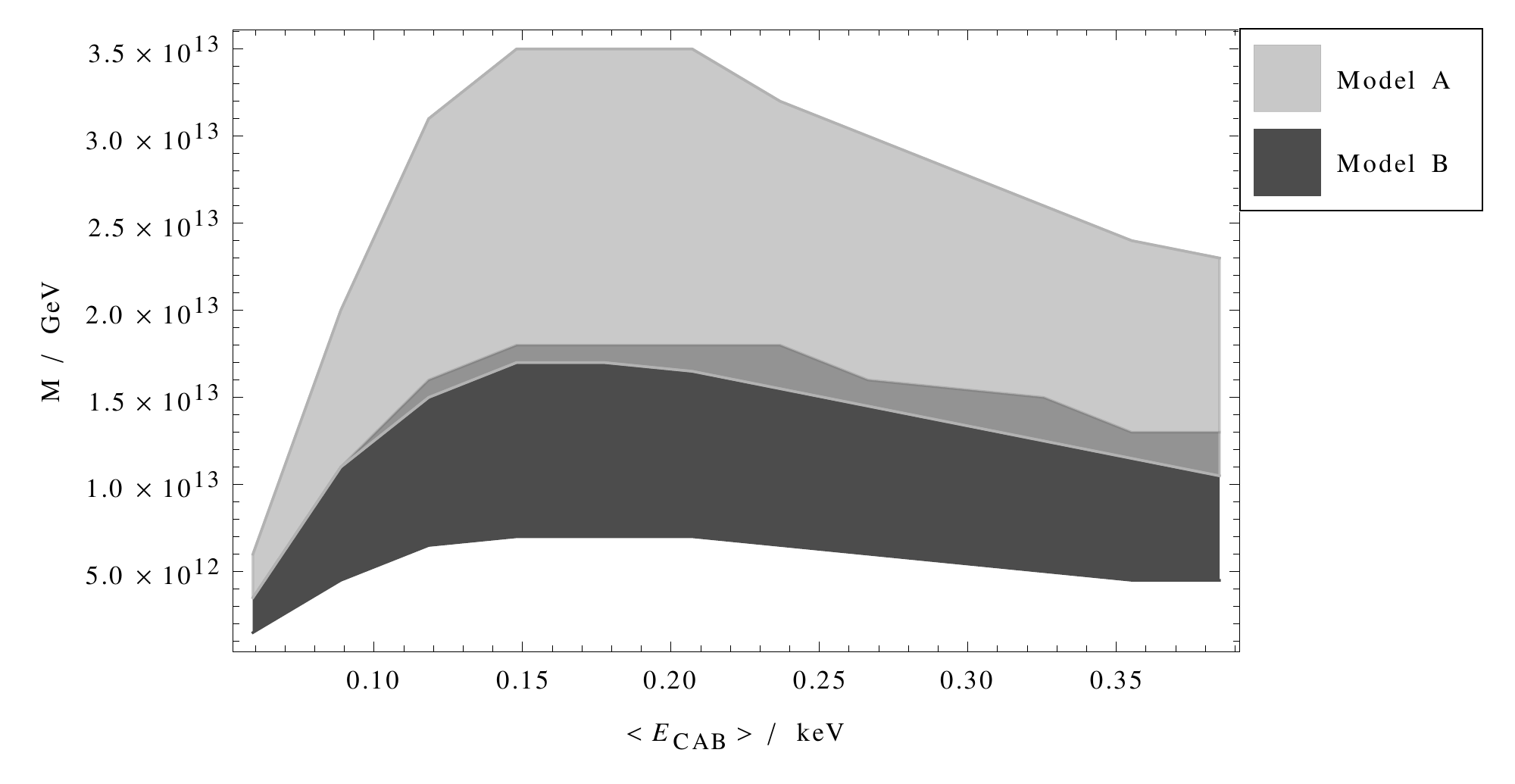}
\caption{Ranges for the inverse ALP-photon coupling $M$ that explain the soft excess in the outskirts of Coma for the two best-fit magnetic field models . Larger $M$s are allowed as then the soft excess is not overproduced. Values for $M$ in the plot are normalized with respect to $\Delta N_{eff}$ as $\frac{M}{\sqrt{\Delta N_{eff} / 0.5}}$.}
\label{boundsM}
\end{figure}

 As spectral information from ROSAT is poor, good fits for the thermal component and the soft excess can be obtained for a range of peak energies of the excess. This continues
 until the CAB component has significant support in the harder ROSAT bands, and would imply a signal beyond the R2 band. This results in a bound of
 $\left\langle E_{CAB} \right\rangle^{\text{max}} \simeq 0.37$\,keV (see Section \ref{softex_sec}). The constraints on $M$ are therefore plotted as a function of
 $\langle E_{CAB} \rangle$ up to $\left\langle E_{CAB} \right\rangle^{\text{max}}$.
  In Figure~\ref{boundsM} we plot the bounds on $M$ for the case where the soft excess luminosity has been extracted by fitting to the CAB spectrum.
   We do not consider $\langle E_{CAB} \rangle < 0.05$ keV since these CAB spectra deposit energy in the R2 band only via their exponential tail, i.e., the peak of the spectrum is far away from the region where the soft excess is observed.

\subsection{Comparison with the excess in the central regions of Coma}
Previous studies of the soft excess~\cite{2002ApJ...576..688B,2013arXiv1312.3947A} used the values for the soft excess luminosity that are extracted from ROSAT count rates assuming the excess is due to a warm thermal component ($kT_{soft}=0.08$\,keV and $A=0.3$).
In order to be compatible with the work done before, particularly the study of the ALP conversion in the central region of Coma~\cite{2013arXiv1312.3947A}, we use this thermal model in the analysis of constraints on $M$.

\begin{figure}[h!]
\centering
\includegraphics[width=\linewidth]{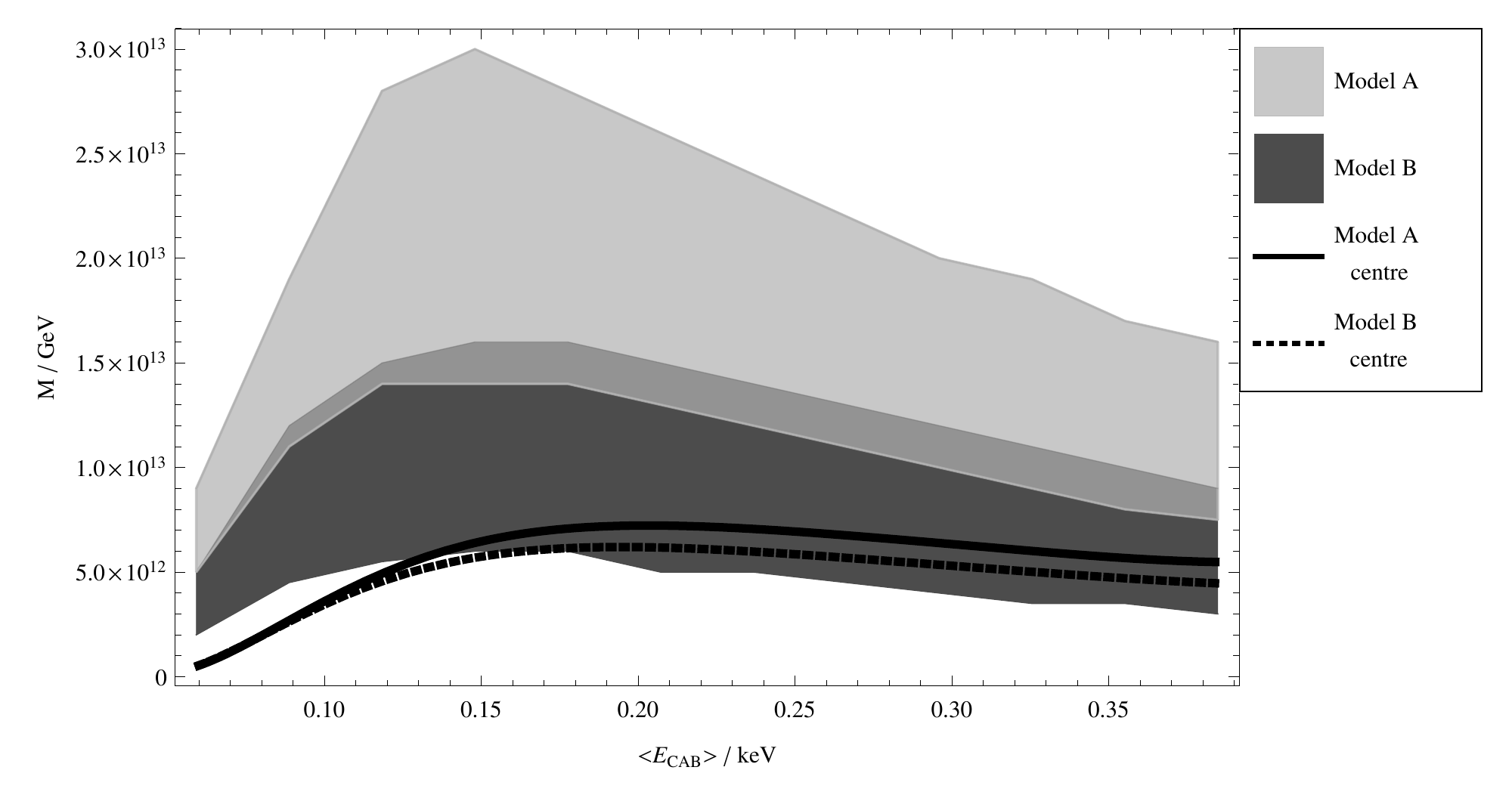}
\caption{Ranges for the inverse ALP-photon coupling $M$ that explain the soft excess for the two best-fit magnetic field models. Here the observed soft excess flux is assumed to be from a warm thermal component (as done in~\cite{2002ApJ...576..688B,2013arXiv1312.3947A}). Larger $M$s are allowed as then the soft excess is not overproduced. Values for $M$ in the plot are normalized with respect to $\Delta N_{eff}$ as $\frac{M}{\sqrt{\Delta N_{eff} / 0.5}}$.}
\label{boundsM_kT}
\end{figure}

For this case we plot the results in Figure \ref{boundsM_kT}. This figure differs from Figure \ref{boundsM} in that the magnitude of the excess luminosity is determined via a fit to a thermal model. While this is not the best approach in the CAB framework, this is necessary for comparison with the analysis in the central region of \cite{2013arXiv1312.3947A},
as this was based by necessity on the magnitudes of the soft excess luminosity determined in \cite{2002ApJ...576..688B} (which involved a fit to a thermal component).

The allowed parameters in the centre are $10^{11} \, \text{GeV} \lesssim M \lesssim 7 \times 10^{12}\,\text{GeV} \sqrt{\Delta N_{eff}/0.5}$ where $M$ is a function of $\left\langle E_{CAB} \right\rangle$ with $50 \, \text{eV} \lesssim \left\langle E_{CAB} \right\rangle \lesssim 250 \,\text{eV}$~\cite{2013arXiv1312.3947A}. The agreement of the values for $M(\left\langle E_{CAB} \right\rangle)$ between the central part of Coma and the outer regions is to within a factor of a few for Model A whereas the agreement when using Model B for the magnetic field is much better (see Figure \ref{boundsM_kT}). Note that Model B also agrees much better with the morphology of the soft excess in the center of Coma than Model A~\cite{2013arXiv1312.3947A}.

Overall, we require slightly higher values for $M$ compared to the central regions. For a fixed $M$, the single domain formula underestimates the luminosity from ALP conversion
compared to a more realistic simulation (see Figure \ref{simvssdM1}). Therefore we expect that in a single domain approach $M$ has to be slightly lowered to agree with the more realistic treatment. Applying this argument to the outer regions of the cluster helps to relax the differences between the constraint on $M$ from the central region and the outer region. Another source of discrepancy is in the details of determining $M$ that explains the amount of soft excess. In the study of the central region~\cite{2013arXiv1312.3947A} the overall predicted luminosity is equated with the observed one to determine $M$. However, in this study we allow for a range of values of $M$ such that the observed flux is within the range of our predictions. The obtained bounds for $M$ therefore contain the specific value for which the overall predicted flux is equal to the overall measured flux making the comparison with the previous studies of the central part of Coma possible.

There are of course also considerable astrophysical uncertainties on the magnetic field and its correlation length at such large distances from the cluster centre. Given this, we find the fact that
the value for $M$ we require differs at most by a factor of a few from the value required in the centre reassuring. This difference is within the reasonable range of uncertainty, and suggests that the CAB explanation of the soft excess can hold consistently in both the central and outskirts region of the cluster.

\section{Conclusions}

This paper has analysed the measured soft X-ray excess in the outer regions ($0.5^\circ-4^\circ$) of the Coma cluster, and has sought to
explain it through conversion of a cosmic axion background  into photons in the cluster magnetic field.
Conversion of such ALPs to photons has been shown in~\cite{2013arXiv1312.3947A} to be able to reproduce the
magnitude and morphology of the soft excess in the central regions ($0^\circ-0.5^\circ$) of the Coma cluster.

We modelled the magnetic field in the outskirts through an extrapolation of the central magnetic field model to large radii.
This clearly involves a significant uncertainty, and
we in part parametrised the uncertainty in the magnetic field through two different models
for the radial dependence of the magnetic field coherence length.

Despite the astrophysical uncertainties and simplified description of ALP propagation, non-trivial ranges for the ALP-photon coupling $M$ and the mean CAB energy $\langle E_{CAB} \rangle$ are obtained. The soft excess in the outer regions can be fitted by a CAB spectrum $ \left\langle E_{CAB} \right\rangle \lesssim 0.37 \text{keV}$ with the range for $M \sim \left(  5 \cdot 10^{12} \,\text{GeV} - 3 \cdot 10^{13}\, \text{GeV}\right) \sqrt{\Delta N_{eff}/0.5}$. These values can at least be interpreted as a lower bound on $M$ since any $M$ greater than that will not overproduce the soft excess and is thus allowed. However, for much larger $M$ one would have to explain the soft excess by a different physical origin than the CAB.

These allowed ranges of $M$ and $\langle E_{CAB} \rangle$ are consistent with the results found in~\cite{2013arXiv1312.3947A} for explaining the magnitude and morphology of the soft excess in the central parts of the Coma cluster (where a value of $M \sim 7 \ti 10^{12} \,\hbox{GeV}$ was found). Hence, the soft excess in the Coma cluster as a whole can be explained by ALP to photon conversion of a CAB.
We have shown that the soft X-ray excess in the outskirts can also be reproduced by a CAB converting to photons, with a value of the
ALP-photon coupling $M$ within at most a factor of a few from that required to reproduce the soft excess in the central regions of Coma.
Given that there is a
significant astrophysical uncertainty in extrapolating the magnetic field measured in the centre of the cluster out into the outskirts,
we think this agreement is acceptable.

An attractive feature of the CAB explanation of the soft excess is therefore that it can allow the same physics to generate the excess in both the central
region and the cluster outskirts. The CAB explanation can be further tested through greater knowledge of the magnetic field, and also by extension to other clusters
other than Coma with information about the
magnetic field, to see if the CAB explanation can reproduce the soft excess found (or not) in these clusters.

\acknowledgments We thank Pedro Alvarez, Stephen Angus, Francesca Day, David Marsh, Alexandre Payez and Andrew Powell for helpful discussions. We are supported by the Royal Society (JC), STFC (DK) and the ERC Starting grant `Supersymmetry Breaking in String Theory' (JC, MR).

\bibliographystyle{JHEP}
\bibliography{coma_outskirts}

\end{document}